\documentclass[a4paper,11pt]{article}
\usepackage{pos}

\title{Looking at the analytic structure of Landau gauge propagators}

\author*[a]{Orlando Oliveira}
\author[a,b]{Alexandre F. Falc\~ao}
\author[a]{Paulo J. Silva}

\affiliation[a]{CFisUC, Department of Physics, University of Coimbra, 3004-516 Coimbra, Portugal}
\affiliation[b]{Department of Physics and Technology, University of Bergen, 5007 Bergen, Norway}

\emailAdd{orlando@uc.pt}
\emailAdd{alfa@uc.pt}
\emailAdd{psilva@uc.pt}

\abstract{We report on a study of the analytical structure of the Landau gauge gluon, ghost and quark  propagators taken from lattice simulations 
using large physical volumes, to better access the IR region, and large gauge ensembles to reduce the statistical uncertainties. The investigation uses 
 Pad\'e approximants to look at poles and branch cuts for each of the propagators. For the gluon propagator we identify complex conjugate poles and a branch point.
 For the ghost propagator the procedure identifies a pole at zero momentum and a branch point for Minkowski-like momenta. The quark propagator appears to
 have a pole for Minkowski-like momenta that is correlated with the pion mass as expected from PCAC.
}

\FullConference{%
 The 38th International Symposium on Lattice Field Theory, LATTICE2021
  26th-30th July, 2021
  Zoom/Gather@Massachusetts Institute of Technology
}

\begin{document}
\maketitle

\section{Introduction and Elements of Pad\'e Approximants}

Lattice simulations are first principles computations that, for the propagators of the fundamental fields, deliver a table of numbers. 
The Landau gauge propagators in QCD computed on the lattice are functions of the Euclidean space momentum. In general,
the interpretation of lattice results is difficult and the computation of  poles and branch cuts in the complex plane  requires going beyond the numerical 
simulations. 
The motivation to look at the analytical structure is multiple as it is connected with confinement,  the generation of bound states, the connection between 
Euclidean and Minkowski momenta, etc.

Our approach to study the analytical structure of the propagators is based on the use of Pad\'e approximants and on Pommerenke’s theorem \cite{Pomm73}.
The use of Pad\'e approximants requires describing the lattice data in Euclidean space by a ratio of polynomials, i.e. a lattice function $\mathcal{G}(p^2)$
computed on a finite number of Euclidean momenta is approximated by
\begin{equation}
   \mathcal{G}(p^2) \approx P^M_N(p^2) = \frac{Q_M(p^2)}{P_N(p^2)} = [M | N] (p^2)
\end{equation}
where  $Q_M$ and $P_N$ are polynomials of order $M$ and $N$, respectively, in $p^2$ and assume that this parameterization of the lattice data
is valid on the all complex plane. Pommerenke’s theorem ensures that for meromorphic functions
the Pad\'e sequences $[M | M +k]$ with fixed $k$ converge in any compact set of the complex plane to 
$\mathcal{G}(p^2)$. In the analysis of the lattice data we consider sequences of Pad\'e approximants $[N| N + 1]$ and look how the poles and zeros 
evolve with $N$. Only those poles whose position does not change with $N$ can be meaningful.

In order to build a Pad\'e approximant to the lattice propagators we use two global optimisation methods, namely Differential Evolution (DE)
and Simulated Annealing (SA), and minimise the function
\begin{equation}
   \sum_j \left( \frac{ \mathcal{G}(p^2_j) - \mathcal{G}_L(p^2_j)  }{\sigma (p^2_j)} \right)^2 
\end{equation}
where the sum runs over the lattice momenta, $\mathcal{G}_L(p^2_j)$ is the lattice estimation of the propagator and
$\sigma (p^2_j)$ the corresponding statistical error.

The details of the analysis of the gluon and ghost propagators that uses data from \cite{Dudal:2018cli} and \cite{Duarte:2016iko}, respectively, can
be found in \cite{Falcao:2020vyr} with the Master thesis \cite{Falcao:2020lxk} giving further details. Herein, we report also on a similar analysis for the quark
propagator using a subset of the results published in \cite{Kizilersu:2021jen}.

\section{Gluon and Ghost Propagators}

Although in \cite{Falcao:2020vyr} we have analysed the lattice data for the gluon and the ghost propagator for various lattice volumes,
herein, due to the lack of space, we only report on the results of the largest physical volume.

\begin{figure*}[htbp]
   \centering
   \includegraphics[scale=0.6]{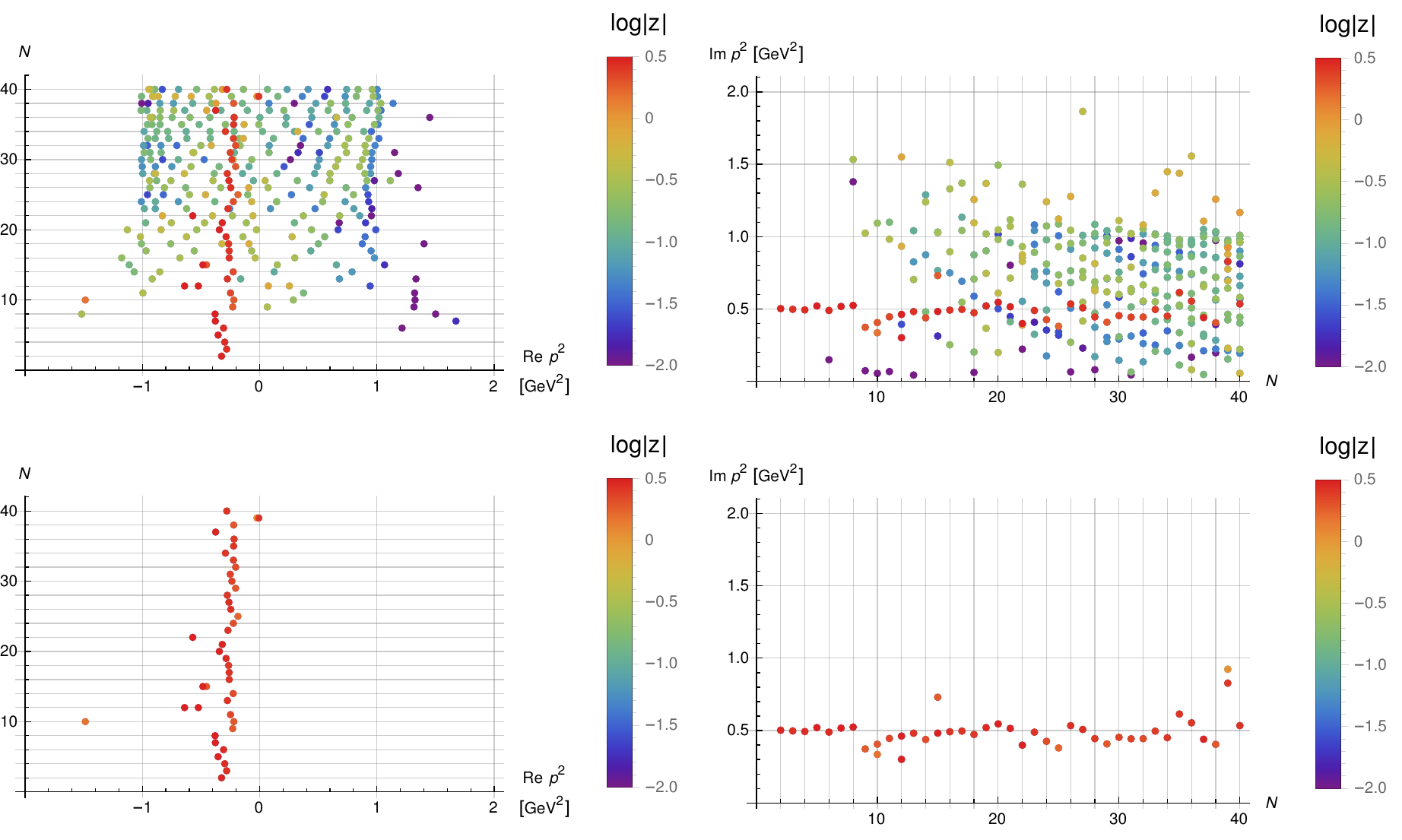} \\
   \includegraphics[scale=0.6]{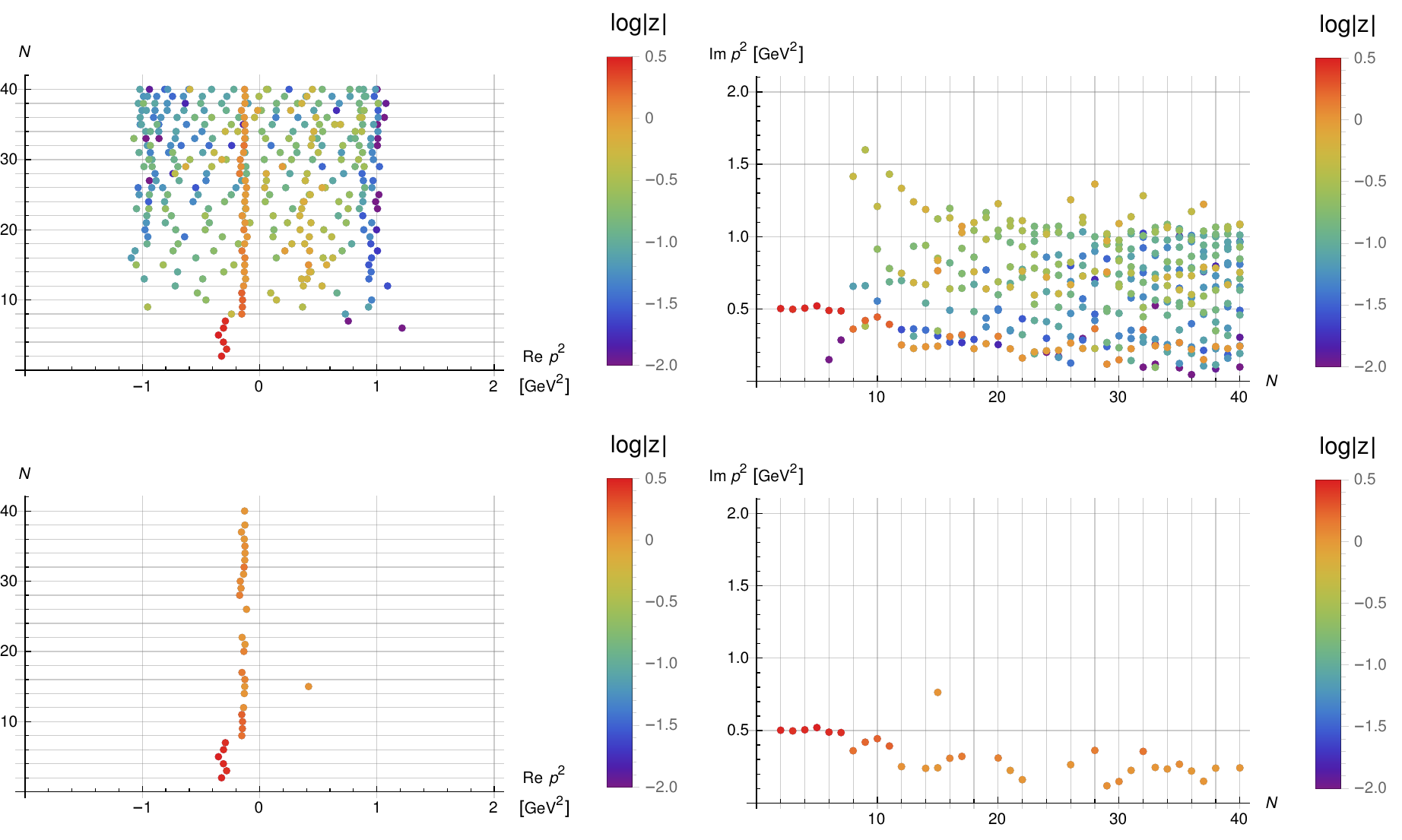} 
   \caption{Sequence of  poles as given by $[N|N+1]$ Pad\'e approximants for the $128^4$ gluon propagator data for off-axis momenta. The upper two plots
   are the results computed with the DE method, while the lower two plot show the results obtained with the SA method. The colour code refers to
   the absolute value of the residua.}
   \label{fig:gluonoff}
\end{figure*}

\begin{figure}[htbp]
   \centering
   \includegraphics[scale=0.4]{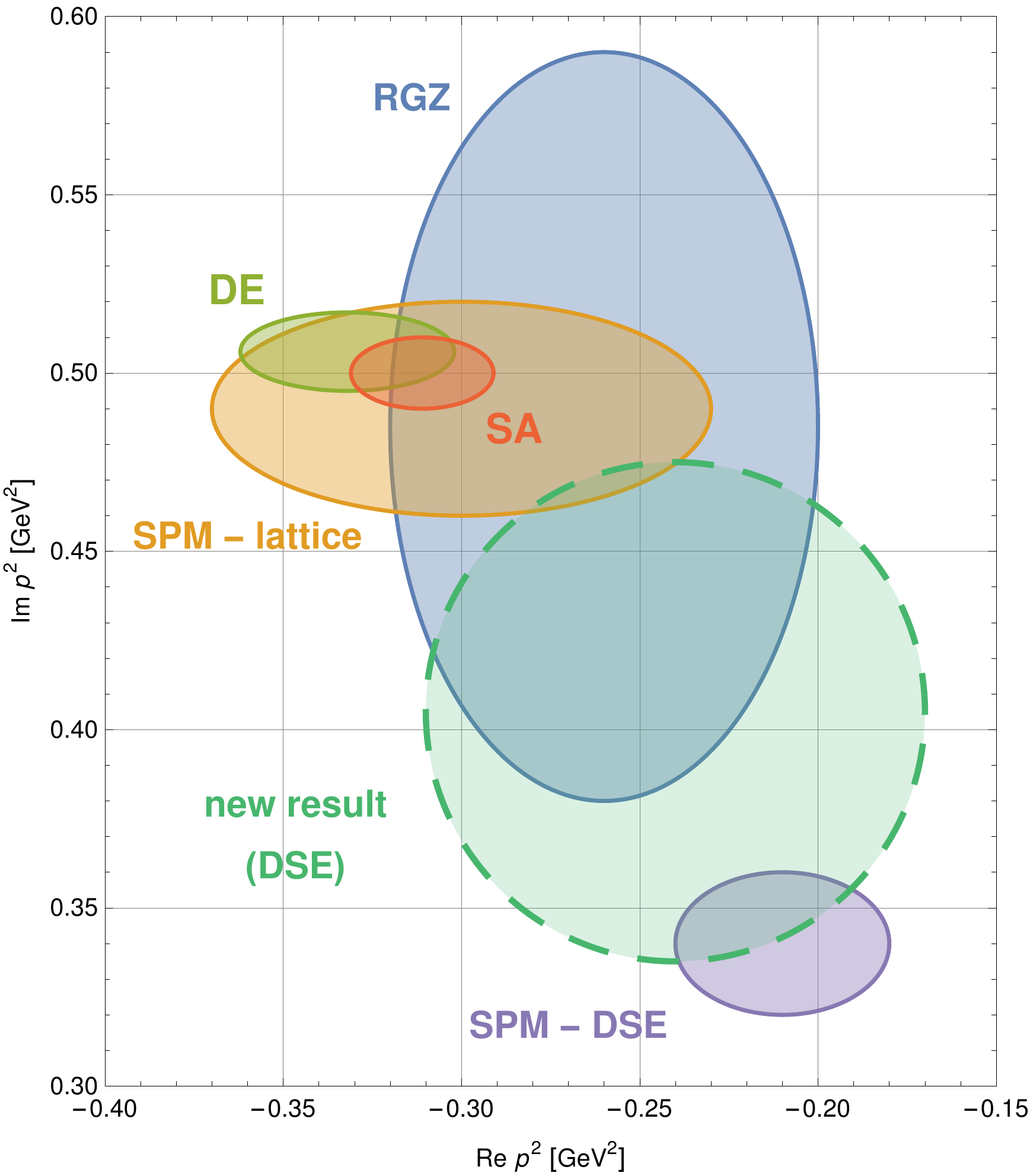} 
   \caption{Estimation of the gluon complex poles using different methods: DE and SA estimations (see text for details); RGZ are the estimations of
   the refined Gribov-Zwanziger analysis of the lattice gluon propagator \cite{Dudal:2018cli}; SPM-Lattice and SPM-DSE are the estimations of
   \cite{Binosi:2019ecz}  using either the lattice data or the propagator from solving the gluon and ghost Dyson-Schwinger equations for pure Yang-Mills SU(3) 
   theory, respectively;
   ``new result (DSE)'' are the results of \cite{Fischer:2020xnb}.}
   \label{fig:GluonComplexPoles}
\end{figure}

\begin{figure*}[htbp]
   \centering
   \includegraphics[scale=0.7]{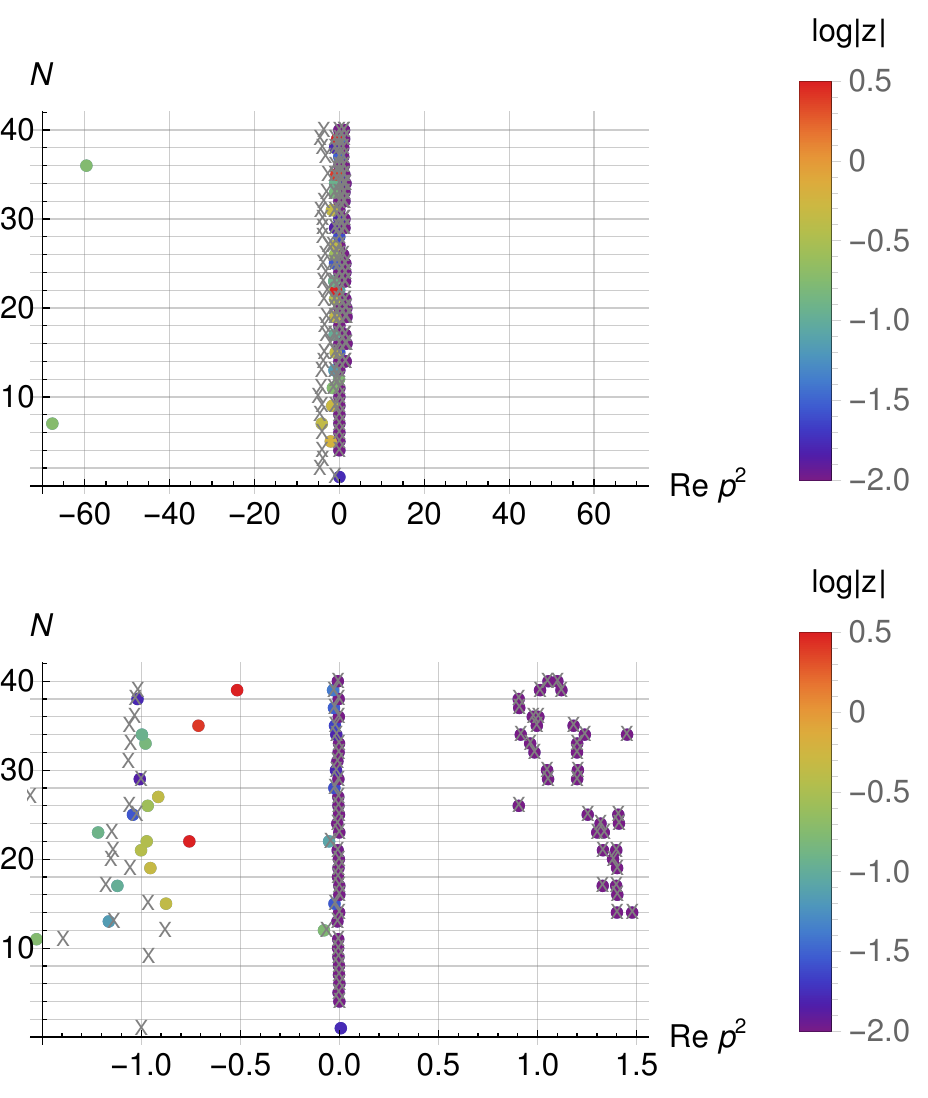} ~
   \includegraphics[scale=0.7]{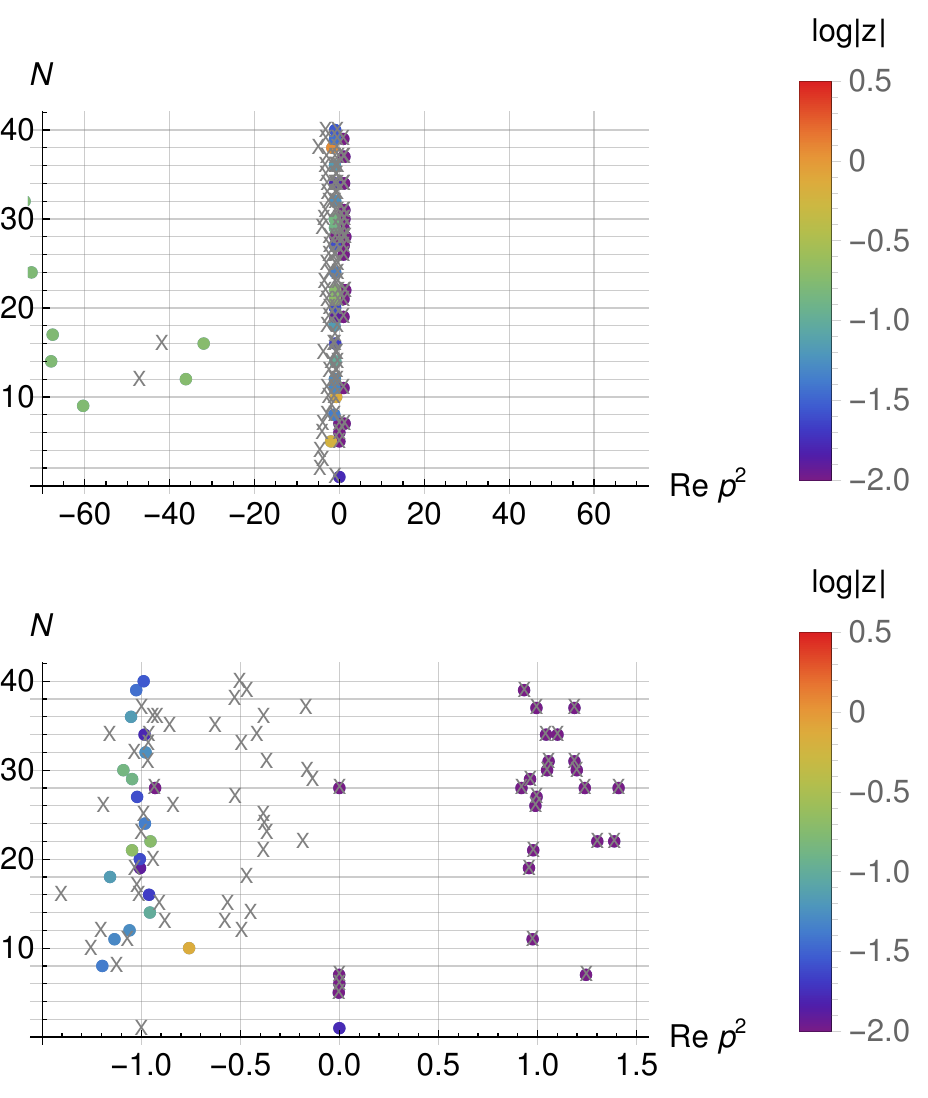} 
   \caption{Sequence of poles (circles) and zeros (crosses) as given by $[N|N+1]$ Pad\'e approximants for the $128^4$ gluon propagator data for on-axis momenta. 
   The left plot are the results computed with the DE method, while the right plot shows the results computed with the SA method. The bottom plots are zooms of
    the region near the origin of the complex plane. The colour code refers to the absolute value of the residua.}
   \label{fig:gluonon}
\end{figure*}

In Fig. \ref{fig:gluonoff} we show  the sequence of poles, computed with the two optimisation methods,
as a function of $N$ for $[N|N+1]$ Pad\'e approximants and for the $128^4$ gluon propagator data for off-axis momenta. As seen, the dominant poles, i.e.
those with higher residuum (in red), are stable and appear at relatively small $N$. Moreover, their positions in the complex plane
as given by the DE or the SA methods is essentially the same. We take this result as an indication that the gluon has complex conjugate poles and, from a selection of
the data in Fig. \ref{fig:gluonoff}, it comes that the poles\footnote{Note that the numbers quoted below do not match exactly those represented in Fig.  
\ref{fig:GluonComplexPoles} that use all the information produced by the Pad\'e analysis \cite{Falcao:2020vyr,Falcao:2020lxk}.} are at
\begin{equation}
   p^2 = \left\{ \begin{array}{lcl}
   - ( 0.343 - 0.220) \pm i ( 0.301 - 0.546)  \text{ GeV}^2, & \mbox{  }  & \text{DE method} ,  \\
   - ( 0.220 - 0.150) \pm i ( 0.227 - 0.444)  \text{ GeV}^2, & \mbox{  }  & \text{SA method}  .
    \end{array} \right.
\end{equation}
In Fig. \ref{fig:GluonComplexPoles} we collect the results of several investigations of the Landau gauge gluon propagator where complex conjugate poles are
observed. We note the good agreement from the different studies.

In Fig. \ref{fig:gluonon} we show the zeros (crosses) and poles (circles) as given by the Pad\'e approximants for on-axis momenta. The two methods do
not give compatible results, with the SA method resulting in a collection of zeros for Minkowski type of momenta that suggests the presence of
a branch cut. The corresponding branch point occurs at the Euclidean momenta $p^2 \sim -0.5$ GeV$^2$ or is closer to the origin of the complex plane.

\begin{figure*}[htbp]
   \centering
   \includegraphics[scale=0.7]{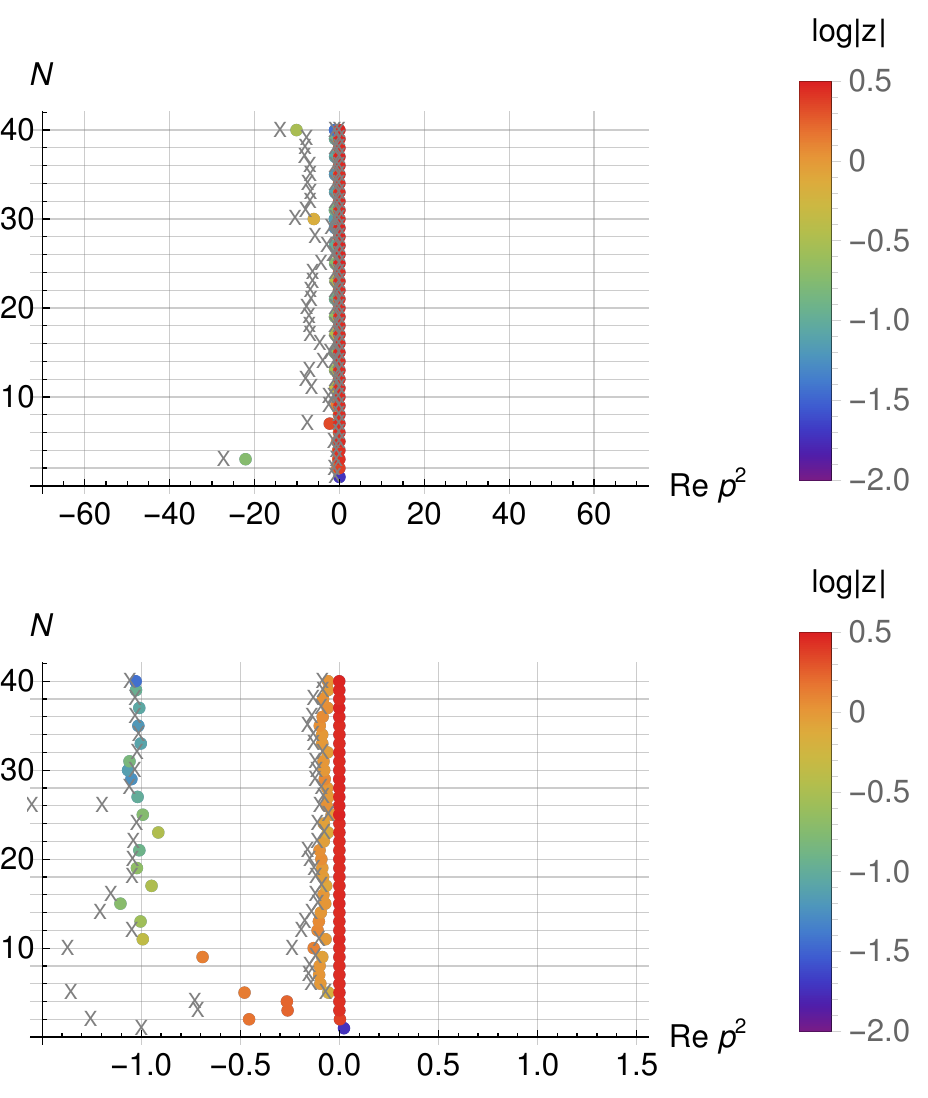} ~
   \includegraphics[scale=0.7]{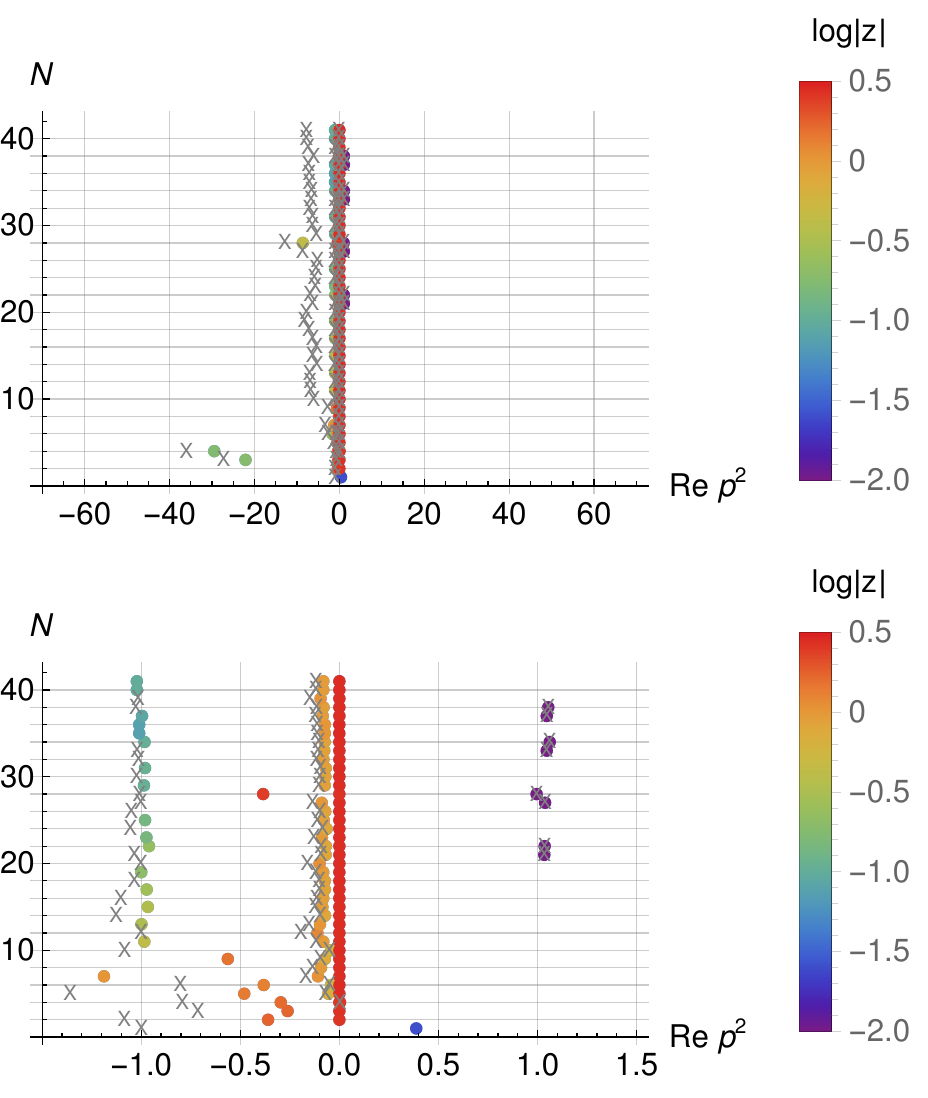} 
   \caption{Sequence of poles (circles) and zeros (crosses) as given by $[N|N+1]$ Pad\'e approximants for the $80^4$ ghost propagator data for on-axis momenta. 
   The left plot are the results computed with the DE method, while the right plot shows the results computed with the SA method. The bottom plots are zooms  
   of the region near the origin of the complex plane. The colour code refers to the absolute value of the residua.}
   \label{fig:ghoston}
\end{figure*}

The same analysis of the ghost data shows no complex poles; see \cite{Falcao:2020vyr} for details. However, for on-axis momenta, see Fig. \ref{fig:ghoston},
the Pad\'e approximants identify a pole at the origin of the complex plane that has the highest absolute value of the residuum, meaning that the ghost propagator 
is of type $Z(p^2)/p^2$. Besides this pole the Pad\'e analysis shows a sequence of zeros that suggests also the presence of a branch cut, with 
the corresponding branch point being at $p^2 \sim - 0.1$ GeV$^2$.

\section{Quark Propagator}

\begin{figure*}[htbp]
   \centering
   \includegraphics[scale=0.26]{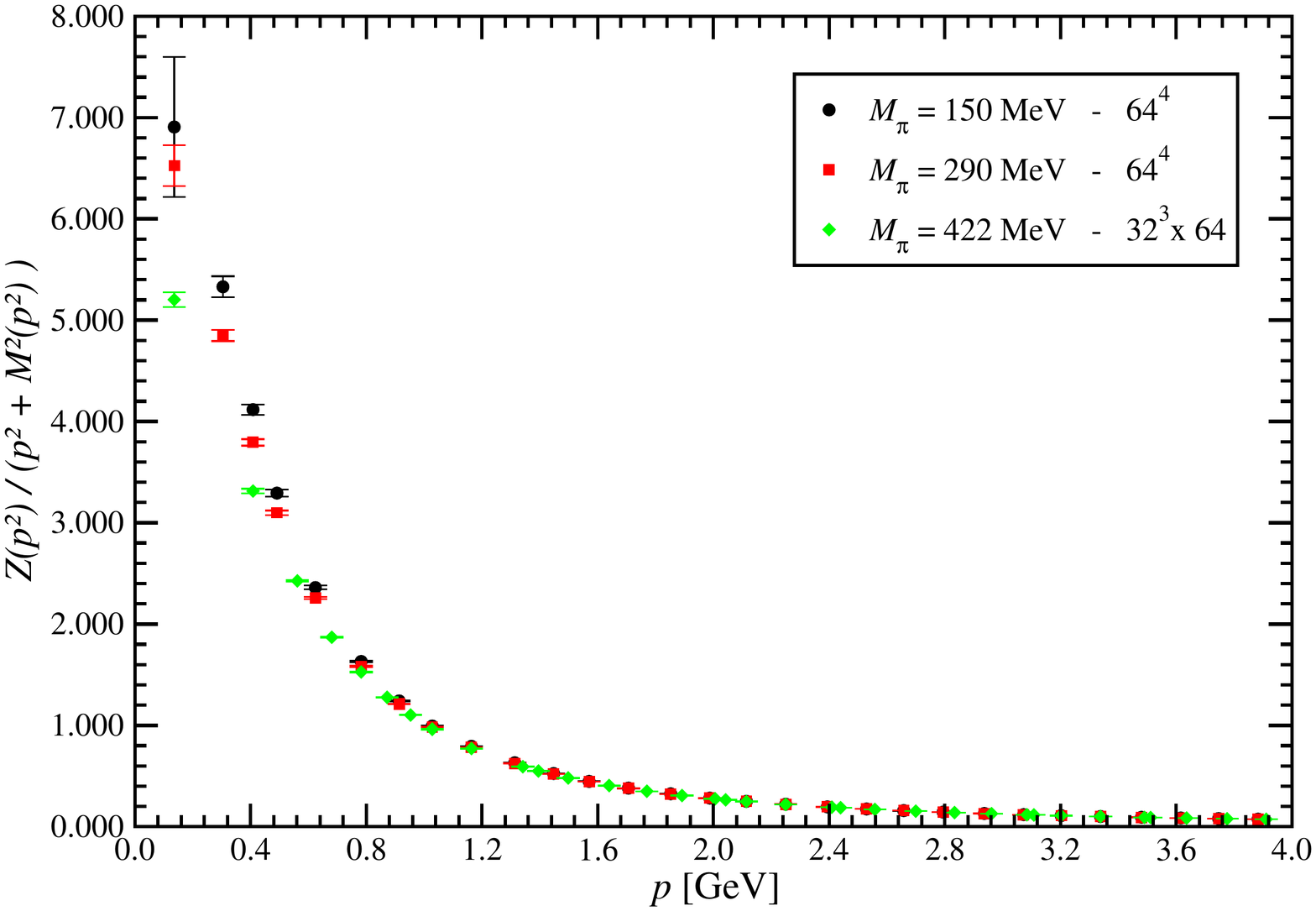} ~
   \includegraphics[scale=0.26]{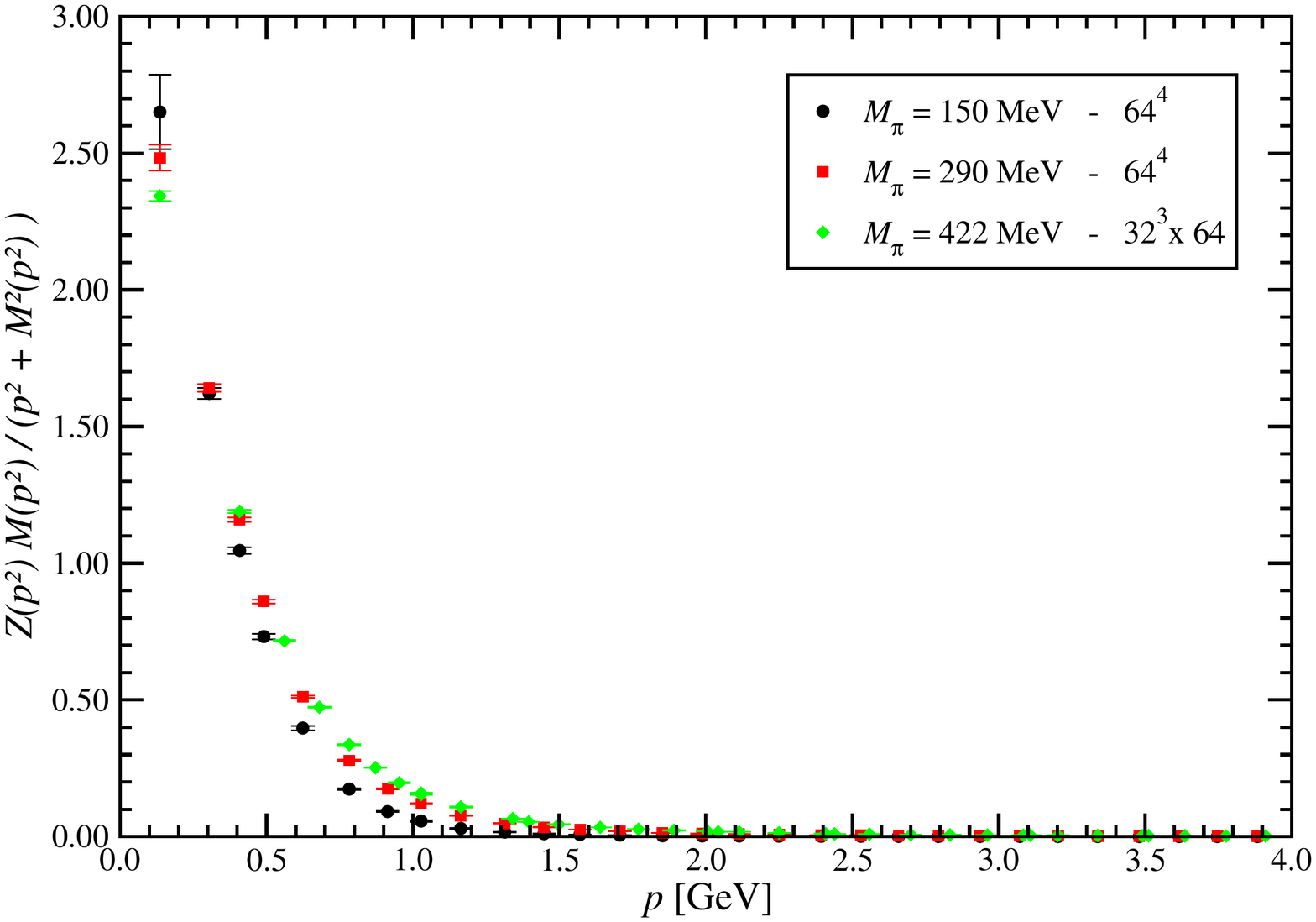} 
   \caption{The quark propagator.}
   \label{fig:quarkprop}
\end{figure*}

\begin{figure*}[htbp]
   \centering
   \includegraphics[scale=0.4]{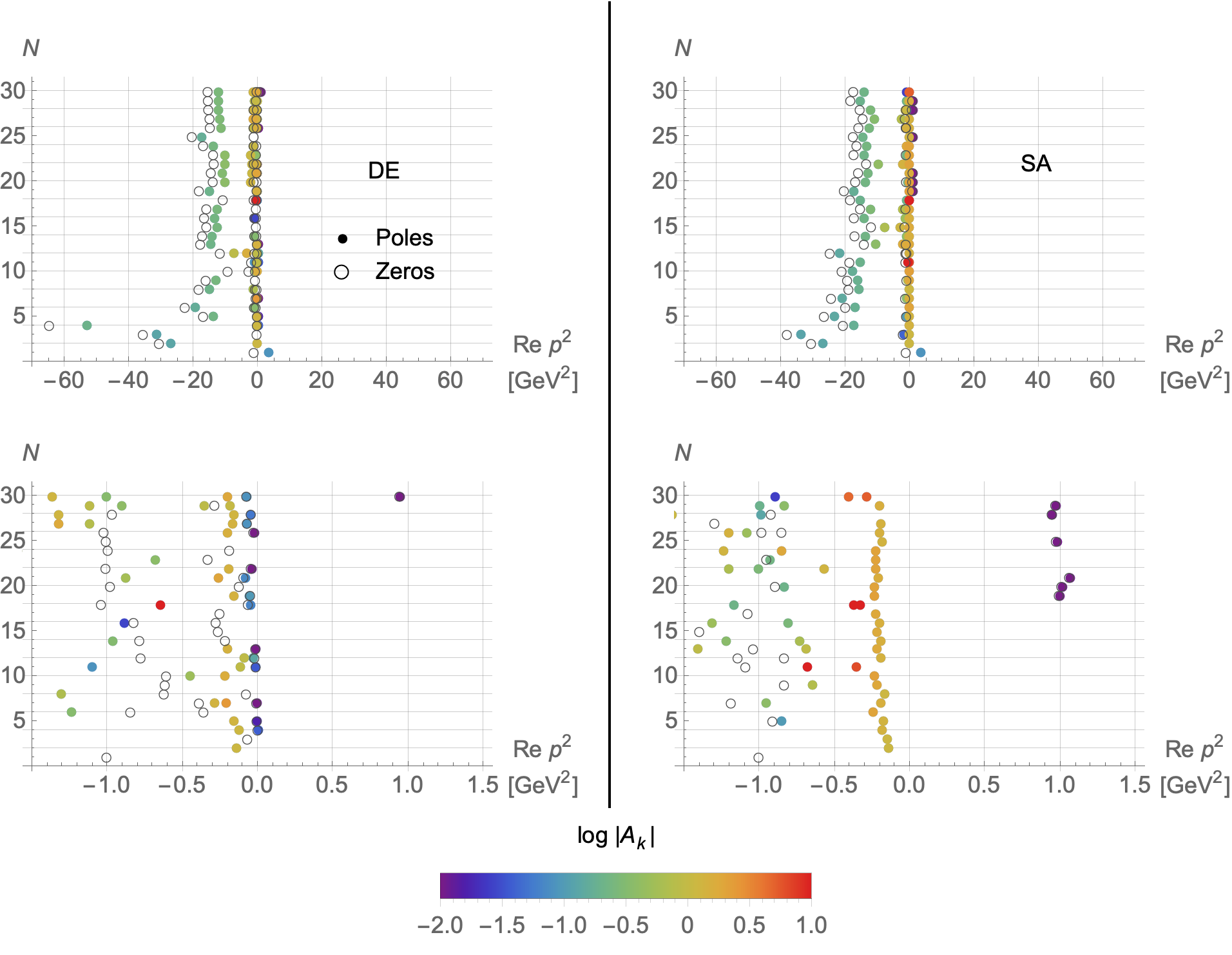} \\
   \includegraphics[scale=0.4]{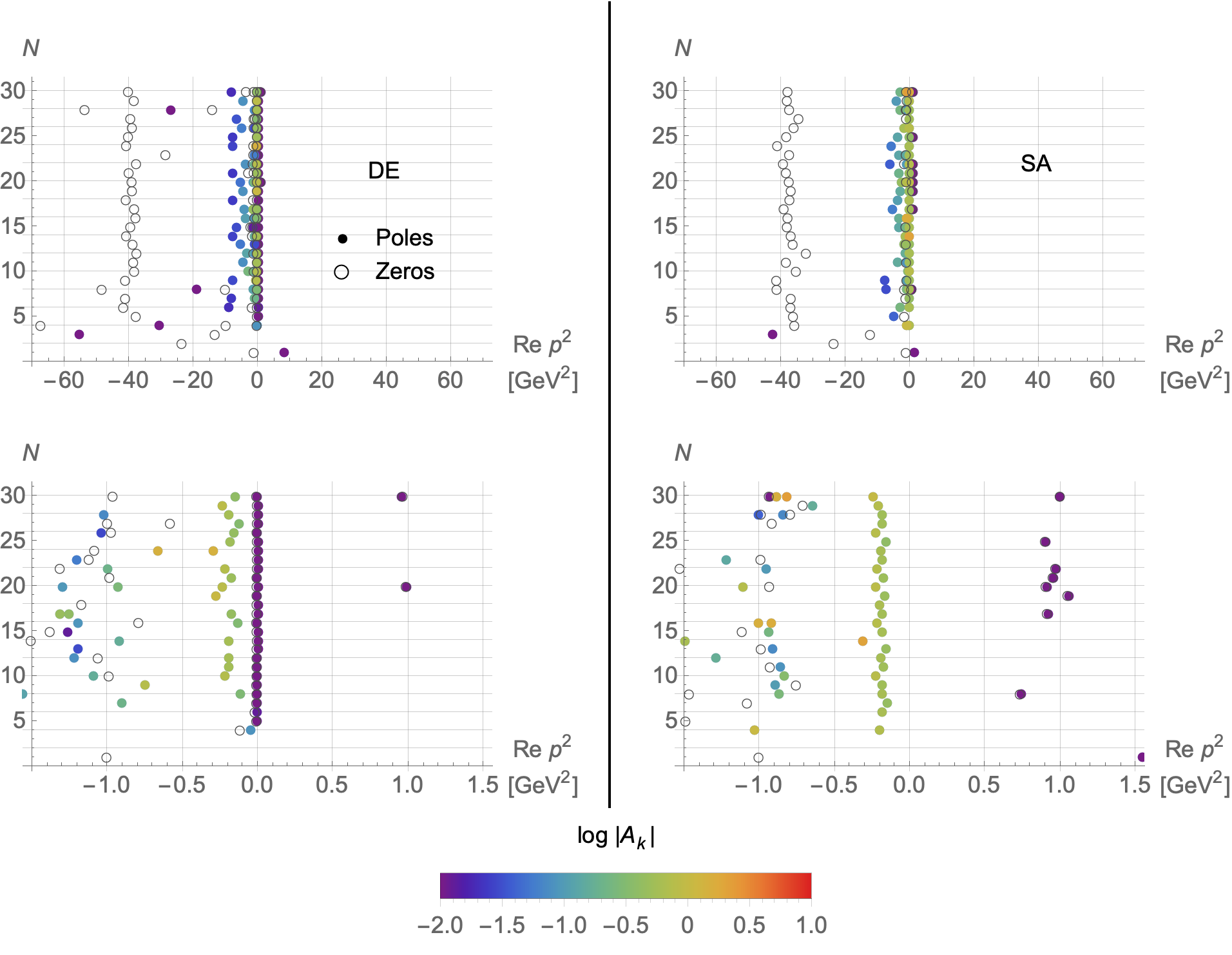} 
   \caption{Poles (full circles) and zeros (open circles) from the analysis with Pad\'e approximants for the quark propagator associated with the ensemble with
   a pion mass of 290 MeV. In the upper plot we show the sequence poles and zeros for the term that multiplies $p \!\!\! /$ and in the bottom plot we show the Dirac scalar
   term.}
   \label{fig:quarkproppoles}
\end{figure*}

For the analysis of the quark propagator we consider the subset of the ensembles generated in \cite{Kizilersu:2021jen} with lattice spacing $a = 0.071$ fm.
Of these data we take for analysis only the simulations with a pion mass of 422 MeV, 290 MeV and 150 MeV. The quark propagator is diagonal in colour space
and its Dirac structure reads
\begin{equation}
   S(p^2) = Z(p^2) \, \frac{ p \!\!\! / + M(p^2)}{p^2 + M^2(p^2)} \ .
\end{equation}   
Fig. \ref{fig:quarkprop} shows the two reconstructed functions associated with the different Dirac structures for the ensembles. 

The Pad\'e analysis of the vector and scalar Dirac functions show no complex poles for any of these propagators. However, surprisingly, it points towards poles
on the negative real axis, i.e. for Minkowski type of momenta. To illustrate the results of the Pad\'e approximant analysis, in Fig. \ref{fig:quarkproppoles} we
report on the results for the quark propagator associated a pion mass of 290 MeV. A stable pole associated with $p^2 \sim - 0.2$ GeV$^2$ is 
clearly observed.  The outcome of the analysis of the three quark propagators points to poles at
\begin{center}
   \centering
   \begin{tabular}{l @{\hspace{0.75cm}} l @{\hspace{0.75cm}} l @{\hspace{0.75cm}} l @{\hspace{0.75cm}} } 
      $M_\pi$           &  150 MeV & 290 MeV&  422 MeV \\
      \hline
    Vector  & 0.19(6) &  0.22(3) & 0.26(3) \\
    Scalar  & 0.16(3) &   0.19(6) & 0.21(6)    
   \end{tabular}
\end{center}
where the pole positions are given in GeV$^2$; recall that the poles appear for Minkowski-like momentum. 
From our results we make the following observations. The pole position obtained from the Dirac vector and the Dirac
scalar functions are consistent within errors. The pole mass correlates with the pion mass and increases when the pion mass increases. Moreover,
the results also show that the pole mass is compatible with $M^2_\pi \propto m_{\text{pole}}$ behaviour as predicted by PCAC. 
This results should be read with care due to the large statistical errors on the pole mass that allow for different types of power law. Finally, the computed
mass poles are of the same order of magnitude as the constituent quark mass used in many quark models.

\section{Summary and Conclusions}

Pad\'e approximants allow to access the analytic structure of the Landau gauge lattice propagators. Accordingly, the gluon propagator is described by
a pair of complex conjugate poles and has a branch cut. The ghost propagator has a pole at zero momentum and a branch cut.
The quark propagator has a pole at Minkowski momenta that grows (linearly) with the pion mass squared. A possible branch cut at $p < 1$ GeV
was also identified.

We plan to improve the precise location of the branch cut for the bosonic propagators and to investigate the presence of multiple poles.

\section*{Acknowledgments}

This work was supported by national funds from FCT -- Fundação para a Ciência e a Tecnologia, I.P., Portugal, within projects UID/FIS/04564/2019,
UID/FIS/04564/2020 and CERN/FIS-COM/0029/2017. A.F.F. acknowledges financial support from FCT (Portugal) under the Project No. UIDB/04564/2020.
P. J. S. acknowledges financial support from FCT (Portugal) under contract CEECIND/00488/2017. 


\end{document}